\documentclass[%
reprint,
superscriptaddress,
amsmath,amssymb,
aps,
prl,
floatfix,
]{revtex4-2}

\usepackage{graphicx}
\usepackage{dcolumn}
\usepackage{bm}
\usepackage{siunitx}
\usepackage[version=4]{mhchem}
\usepackage{mathtools}
\usepackage{tikz}
\usepackage{xcolor} 
\usepackage[normalem]{ulem}
\usepackage{hyperref}

\DeclareSIUnit\angstrom{\text {Å}}

\newcommand{\vect}{\mathbf}

\newcommand{\phiE}{\phi_{\mathrm{E}}}

\preprint{APS/123-QED}
\begin{document}

\title{Machine Learning the Energetics of Electrified Solid/Liquid Interfaces}

\author{Nicolas Bergmann}
\affiliation{%
 Fritz-Haber-Institut der Max-Planck-Gesellschaft,\\
 Faradayweg 4-6, D-14195 Berlin, Germany
}
\author{Nic\'{e}phore Bonnet}
\affiliation{%
 Laboratory for Theory and Simulation of Materials, EPFL,\\
 Lausanne
}
\author{Nicola Marzari}
\affiliation{%
 Laboratory for Theory and Simulation of Materials, EPFL,\\
 Lausanne
}
\author{Karsten Reuter}
\affiliation{%
 Fritz-Haber-Institut der Max-Planck-Gesellschaft,\\
 Faradayweg 4-6, D-14195 Berlin, Germany
}
\author{Nicolas G. H\"{o}rmann}%
\email{hoermann@fhi.mpg.de}
\affiliation{%
 Fritz-Haber-Institut der Max-Planck-Gesellschaft,\\
 Faradayweg 4-6, D-14195 Berlin, Germany
}

\begin{abstract}
We present a response-augmented machine learning (ML) approach to the energetics of electrified metal surfaces. We leverage local descriptors to learn the work function as the first-order energy change to introduced bias charges and stabilize this learning through Born effective charges. This permits the efficient extension of ML interatomic potential architectures to include finite bias effects up to second-order. Application to OH at Cu(100) rationalizes the experimentally observed pH-dependence of the preferred adsorption site in terms of a non-Nernstian charge-induced site switching.
\end{abstract}

\maketitle
Predictive-quality first-principles modeling and simulation has become indispensable for studying electrochemical interfaces. By accessing the detailed atomic structure and prevailing interactions, it provides deep mechanistic insight and generates ideas for the design of improved electrocatalysts \cite{norskov_origin_2004, resasco_2022_enhancing, gross_2022_abinitio, salanne_2022_microscopic}. Compared to its analog usage at solid/gas interfaces and thermal catalysis, first-principles modeling of electrified solid-liquid interfaces nevertheless still lags behind \cite{schwarz_electrochemical_2020}, largely because of difficulties to describe the extended double layer (DL) that builds up at biased conditions. Even within common approximations like implicit solvation \cite{dabo_firstprinciples_2010, dabo_abinitio_2010, nattino_continuum_2018, ringe_implicit_2022}, the description of complex and potential-dependent (near-surface) dynamics remains a challenging, computational burden. 

In this respect, application of machine-learning interatomic potentials (MLIPs) \cite{behler_four_2021,deringer_gaussian_2021, jacobs_practical_2025} as fast surrogates to the first-principles calculations is highly appealing. MLIPS access longer time and larger length scales, a prerequisite to derive reliable macroscopic insights from the atomistic interfacial structures and reactions \cite{raman_insights_2024, stocker_estimating_2023, timmermann_iro2_2020, bruix_firstprinciples_2019}. Notwithstanding, the necessity to accurately capture the local fields that build up at the electrified interface prevents the straightforward usage of established short-range MLIP architectures \cite{omranpour_machine_2025}. To this end, an important strand of ongoing developments centers on incorporating long-range electrostatic interactions \cite{grisafi_incorporating_2019, ko_fourth-generation_2021, gao_self-consistent_2022, staacke_kernel_2022, vondrak_qpac_2023, grisafi_predicting_2023, hisama_molecular_2024, monacelli_electrostatic_2024}. While this has led to a successful modeling of the atomistic structure of the interface at the potential of zero charge \cite{zhang_molecular-scale_2024}, there are still challenges at applied electrode potential conditions. For instance, underlying charge equilibration schemes can suffer from unphysical, partially metallic behavior of the electrolyte, while the need to solve for electrostatic potentials and atomic charges self-consistently generally increases the computational costs, thereby reducing the very efficiency gain sought for with MLIPs.

As such, a complementary strand of developments appears promising: Models incorporating the effects of electric fields by machine learning relevant response effects \cite{christensen_operators_2019, gastegger_machine_2021, gigli_thermodynamics_2022, joll_machine_2024, falletta_unified_2025, schmiedmayer_derivative_2024}. With previous works addressing isolated molecules, bulk solids or liquid water, we here transfer this concept to electrified interfaces formed between a liquid electrolyte and a metallic electrode. We do this in the understanding that the explicit account of local electrostatic fields generated by electronic bias charges is crucial for modeling adsorbed species and solvent molecules in the immediate interfacial layer(s) \cite{patel_non-Nernstian_2025, dudzinski_first_2023, bergmann_ab_2023, beinlich_field_2022, hormann_converging_2024, li_deciphering_2025, bonnet_chemisorbed_2014}. In contrast, knowledge about water and ion distributions in the further (diffuse) DL is predominantly important for appropriate referencing to experimental conditions, i.e. relating local bias charges $q$ to the applied electrode potential $\phiE$.

In this spirit, we present the "Response Analysis in $z$-ORientation" (RAZOR) model, which machine-learns the first-order derivatives of the potential energy and forces for interfacial species at extended interfaces with respect to $q$, while an appropriate referencing is achieved by effectively describing the extended DL part at the level of an implicit solvent model. By treating $q$ as a perturbation to the charge-free energy, RAZOR offers a robust and computationally efficient extension of common MLIPs to the case of finite biases up to second-order. Using OH adsorbed on Cu(100) as a prototypical showcase, we demonstrate that RAZOR-MLIP based molecular dynamics (MD) simulations at applied bias reliably describe non-Nernstian, interfacial behavior and reveal a potential-dependent switching of the adsorption site. 

\begin{figure*}[htbp]
    \centering
    \includegraphics[scale=1]{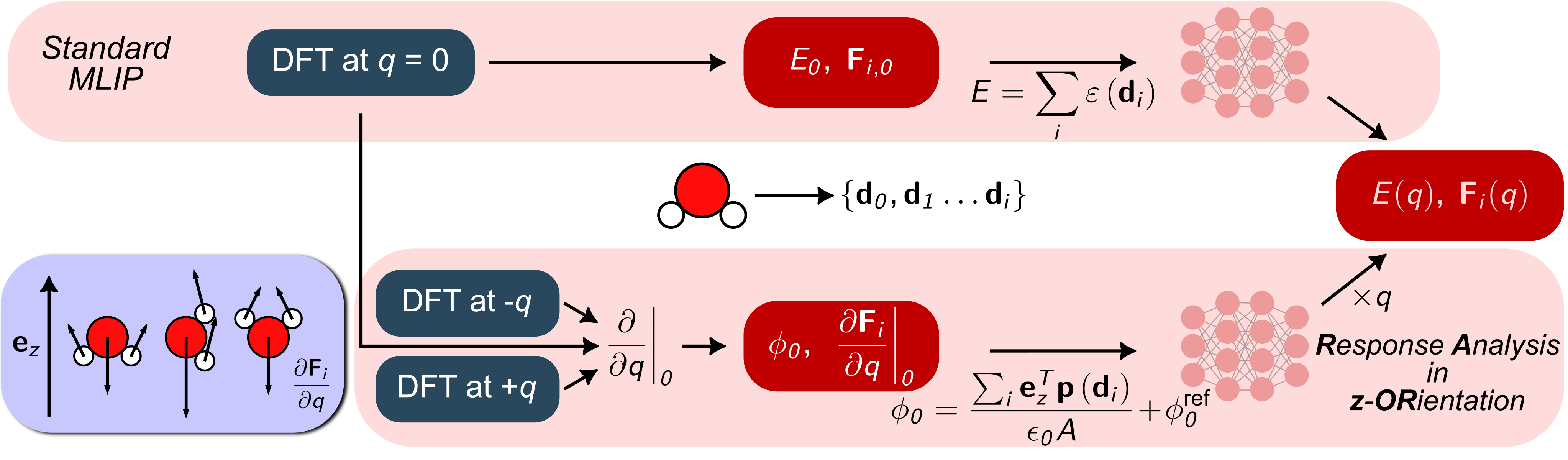}
    \caption{Schematic workflow showing how combining the "Response Analysis in $z$-ORientation" (RAZOR) approach with a standard MLIP provides the energy $E^\alpha$ and atomic forces $\vect{F}_{i}$ of an interfacial configuration $\alpha$ at excess bias charge $q$. RAZOR machine-learns the energy change associated with the response parameter $\phi^\alpha_{0} = \partial E^\alpha / \partial q$ at $q=0$, stabilized by including $- \partial \phi^\alpha_{0} / \partial \vect{r}_{i} = \partial \vect{F}_{i} / \partial q$, equivalent to Born effective charges $\vect{Z}_{i}^{*}$ (inset on the left).}
    \label{fig:NN_workflow}
\end{figure*}
The general objective targeted here is to derive stabilities of interfacial states as a function of $\phiE$, measured relative to a universal potential reference. Practical calculations largely address this in half-cell setups, where only a single electrode and the DL is present. The stability of an interface configuration $\alpha$ is then described by the electronically grand-canonical free energy $\Omega^{\alpha}(\phiE)$, where $\alpha$ summarizes the configuration's chemical composition $Z_i$ and atomic positions $\vect{r}_i$. Constant-potential conditions can be realized via introduction of a potentiostat \cite{bonnet_first-principles_2012, chen_atomistic_2023}, or by Legendre-transforming electronically canonical free energies $F^{\alpha}(q)$ at constant $q$ \cite{beinlich_controlled_2023, hormann_converging_2024, kastlunger_thermodynamic_2024}.

Focusing on the latter route, we Taylor expand the $q$-dependence of the total-energy component of $F^{\alpha}(q)$ around $q=0$ (subscript 0),
\begin{eqnarray}
    \nonumber 
    E^{\alpha}(q) &=& E_{0}^{\alpha} \;+\; \left.\frac{\partial E^{\alpha}}{\partial q}\right|_{0} q \;+\; \frac{1}{2}\left.\frac{\partial^{2}E^{\alpha}}{\partial q^{2}}\right|_{0} q^{2} \;+\; \dots \\
    &=& E_{0}^{\alpha} \;+\; \phi^{\alpha}_{0} q \;+\; \frac{1}{2}\frac{1}{C_{\mathrm{el}, 0}^{\alpha}} q^{2} \;+\; \dots ~ .
    \label{eq:energy_charge_taylor}
\end{eqnarray}
Here, $\phi^{\alpha}_{0}$ is the work function (divided by the electronic charge $e$) of $\alpha$ and $C^{\alpha}_{\mathrm{el}, 0}$ is the second-order response parameter, both at $q=0$. Here, we refer to the latter as the "electronic capacitance" $C^{\alpha}_{\mathrm{el}, 0}$ (cf. Refs. \onlinecite{bonnet_chemisorbed_2014, binninger_piecewise_2021, beinlich_controlled_2023, beinlich_theoretical_2023} and the discussion therein). The force on any atom $i$ in $\alpha$ at $q$ is correspondingly
\begin{eqnarray}
    \nonumber
    \vect{F}_{i}(q) 
        &=& -\frac{\partial E^{\alpha}(q)}{\partial \vect{r}_{i}} \\
        &=&   \vect{F}_{0, i}
            - \frac{\partial\phi_{0}^{\alpha}}{\partial\vect{r}_{i}} q 
            - \frac{1}{2}\frac{\partial \left(C^{\alpha}_{\mathrm{el},0}\right)^{-1}}{\partial \vect{r}_{i}} q^{2}
            + \dots \quad ,
    \label{eq:forces_charge_taylor}
\end{eqnarray}
which clarifies that machine-learning the three contributions $E^{\alpha}_{0}$, $\phi^{\alpha}_{0}$, and $C^{\alpha}_{\mathrm{el}, 0}$ (all at $q=0$), as well as their respective $\vect{r}_{i}$ derivatives is necessary to build a second-order correct surrogate model for $E^{\alpha}(q)$.

Many MLIP architectures exist to learn $E_{0}^{\alpha}$ and $\vect{F}_{0, i}$ in field-free environments. Second generation MLIPs express the total energy $E^{\alpha}_{0}$ as the sum of contributions $\varepsilon(\vect{d}_{i})$ from every atom $i$ in the system \cite{behler_four_2021, bartok_gaussian_2010, smith_ani_2017, batatia_mace_2022, batzner_e3-equivariant_2022, zhang_deep_2022},
\begin{equation}
    E^{\alpha}_{0} 
    = \sum_{i \in \alpha} \varepsilon_{0, i} 
    = \sum_{i \in \alpha} \varepsilon_{0}(\vect{d}_i)
    \quad .
    \label{eq:ML_E_input}
\end{equation}
These contributions depend on the local atomic environment of the respective atom as encoded into a descriptor vector $\vect{d}_{i}$, typically respecting fundamental symmetries. Therefore, the actual learning of the $\varepsilon_{0}(\vect{d}_{i})$ requires only first-principles data at $q=0$, and is stabilized by simultaneously learning on the equally provided forces
\begin{equation}
    \vect{F}_{0, i} \;=\; - \frac{\partial E^{\alpha}_{0}}{\partial \vect{r}_{i}}  \quad .
    \label{eq:force_energy_derivative}
\end{equation}

To the best of our knowledge, no machine-learning model exists that immediately predicts work functions, i.e. interfacial potential drops $\phi_{0}^{\alpha}$. However, recent MLIP models have included external electric field response to describe bulk structures and liquid water \cite{christensen_operators_2019, gastegger_machine_2021, gigli_thermodynamics_2022, joll_machine_2024, falletta_unified_2025}, using analogous logics to Eq.~(\ref{eq:energy_charge_taylor}), albeit as a function of the electric field vector $\vect{E}$ instead of $q$. In this case, the first-order terms are the structure's Cartesian polarization vector $\vect{P}^\alpha$ and the Born effective charge rank-two tensor $\vect{\hat{Z}}_{i}^{*}$ \cite{falletta_unified_2025, stocco_electric-field_2025, wang_dynamical_2022} defined by
\begin{equation}
    -\frac{\partial\vect{P}^{\alpha}}{\partial\vect{r}_{i}} 
    = -\frac{\partial^2 E^{\alpha}}{\partial \vect{E} \partial\vect{r}_{i}}
    = -\frac{\partial^2 E^{\alpha}}{\partial\vect{r}_{i} \partial\vect{E}} 
    = \frac{\partial\vect{F}_i}{\partial\vect{E}} 
    = \vect{\hat{Z}}_{i}^{*} \quad .
    \label{eq:Born_effective_charges}
\end{equation}
At extended planar metal electrodes, excess charges yield interfacial fields perpendicular to the surface, suggesting that the broken-symmetry situation at charged interfaces is equivalent to an uniaxial electric field along the $z$-direction with $\vect{E} = E_{z} \vect{e}_{z} = (0,0,E_{z})^{T}$. Then, only the $z$-projected quantities are of relevance: the scalar $P^{\alpha}_{z}$ instead of the polarization vector $\vect{P}^{\alpha}$ and the Cartesian vector $\vect{Z}_{i}^{*}$ instead of the full tensor $\vect{\hat{Z}}_{i}^{*}$.

The intensive potential drop $\phi^{\alpha}_{0}$ can furthermore be connected to the extensive interface polarization $P^{\alpha}_{z}$ (total dipole moment) through the Helmholtz equation \cite{smoluchowski_anisotropy_1941, li_oxygen_2002, ibach_physics_2006, beinlich_field_2022}
\begin{equation}
    P^{\alpha}_{z} 
    = \epsilon_{0} A \Delta \phi_{0}^{\alpha}
    = \epsilon_{0} A \left(\phi^{\alpha}_{0} - \phi^{\mathrm{ref}}_{0}\right)
    \quad .
\label{eq:helmholtz_equation}
\end{equation}
Here $\epsilon_{0}$ is the dielectric constant of vacuum, $A$ the cell's surface area, and $\phi_{0}^{\mathrm{ref}}$ is the work function of a reference state (e.g. the slab without adsorbates). Equation (\ref{eq:helmholtz_equation}) shows that machine-learning the first-order $q$ response is essentially an identical problem as learning the first-order $E_{z}$ response, only with exchanged learning targets
\begin{eqnarray}
    P^{\alpha}_{z} 
    &\leftrightarrow& 
    \epsilon_{0} A \Delta \phi_{0}^{\alpha}
    \label{eq:q_vs_Efield_learning_P} \\
    \vect{Z}_{i}^{*}
    &\leftrightarrow&
    \epsilon_{0} A \frac{\partial\vect{F}_{i}}{\partial q} 
        = 
    -\epsilon_{0} A \frac{\partial \phi^{\alpha}_{0}}{\partial \vect{r}_i}
    \quad .
\label{eq:q_vs_Efield_learning_Z}
\end{eqnarray}
While $\Delta \phi_{0}^{\alpha}$ is available from first-principles calculations at $q=0$, obtaining $\vect{Z}_{i}^{*}$ necessitates calculating $q$-derivatives of $\vect{F}_{i}$. As detailed in the Supplemental Material (SM) (Section SIV.2), we evaluate these numerically by performing calculations at $+q$ and $-q$ for each interfacial structure in the training set. While this formally triples the first-principles computational costs for the training of a RAZOR-MLIP, use of the electron density of the charge-neutral calculation for SCF-initialization reduces this in practice to less than a doubling.

The analogy to Eqs. (\ref{eq:ML_E_input}) and (\ref{eq:force_energy_derivative}) suggests that the extensive quantity $P^{\alpha}_{z}$, cf. Eq. (\ref{eq:q_vs_Efield_learning_P}), can be learned using established MLIP architectures by expanding the $z$-component of the total dipole moment into atomic contributions $p_{z,i}$
\begin{equation}
    P^{\alpha}_z 
    = \sum_{i \in \alpha} p_{z,i} 
    = \sum_{i \in \alpha} p_z({\vect{d}}_i) \quad ,
    \label{eq:dipole_sum_of_local_dipoles}
\end{equation}
stabilized by the atomic contributions $\vect{Z}_{i}^{*}$, cf. Eq. (\ref{eq:q_vs_Efield_learning_Z}). This constitutes a core aspect of RAZOR. In addition to learning charge-dependent energetics, it e.g. equally provides a new pathway to separate global, interfacial potential drops into local contributions without relying on arbitrary electronic charge-density decomposition schemes to define local charges and dipoles.

This leaves the electronic capacitance $C^{\alpha}_{\mathrm{el},0}$ as the last quantity in Eq. (\ref{eq:energy_charge_taylor}) that one needs to machine-learn to establish the RAZOR-MLIP surrogate. $C^{\alpha}_{\mathrm{el},0}$ can be understood from a series circuit of capacitors $(C^{\alpha}_{\mathrm{Q,0}})^{-1} + (C^{\alpha}_{\mathrm{XC,0}})^{-1} + (C^{\alpha}_{\mathrm{DL,0}})^{-1}$, with the three terms being the quantum, the exchange-correlation, and the DL capacitance, respectively \cite{binninger_piecewise_2021, oschinski_constant_2024}. $C^{\alpha}_{\mathrm{Q,0}}$ is related to the spatial variation of the local density of states at the Fermi-level and somewhat analogous to the molecular polarizability. For the metallic electrodes studied here, the contributions of $C^{\alpha}_{\mathrm{Q,0}}$ and $C^{\alpha}_{\mathrm{XC,0}}$ to $C^{\alpha}_{\mathrm{el},0}$ are negligible \cite{binninger_piecewise_2021, oschinski_constant_2024}. Thus, $C^{\alpha}_{\mathrm{el},0}$ is predominantly defined by $C^{\alpha}_{\mathrm{DL,0}}$, which in the present description subsumes all second-order energy contributions due to the collective response of the DL. 

In fully explicit atomistic descriptions, $C^{\alpha}_{\mathrm{el},0}$ ($C^{\alpha}_{\mathrm{DL,0}}$) could be determined by integrating out degrees of freedom that reach deep into the bulk electrolyte, requiring exhaustive sampling \cite{scalfi_semiclassical_2020, ntim_molecular_2023, zhang_molecular-scale_2024}. Within an MLIP approach, such a sampling could be achieved by including a long-range electrostatic description \cite{grisafi_incorporating_2019, ko_fourth-generation_2021, gao_self-consistent_2022, staacke_kernel_2022, grisafi_predicting_2023, hisama_molecular_2024, monacelli_electrostatic_2024}, albeit with increase in computational effort. Here, we instead use an implicit solvent model, which provides an acceptable estimate for the major contributions of the deeper DL \cite{andreussi_electrostatics_2014, kastlunger_controlled-potential_2018, hormann_electrosorption_2020, li_deciphering_2025}. As we already compute different $q$ states to machine-learn $\phi_{0}^{\alpha}$, calculating $C^{\alpha}_{\mathrm{el, 0}}$ brings no additional first-principles computational cost. While $C^{\alpha}_{\mathrm{el},0}$ in implicit solvent models is often rather independent of specifics of interface geometries like adsorption sites, it does depend on the overall interface composition $Z_i$ and termination, e.g. surface coverage $\theta$ \cite{hoermann_thermodynamicCVs_2021, beinlich_field_2022, li_deciphering_2025, bergmann_ab_2023} (see SM, Fig. S8). If we therefore consider $C^{\alpha}_{\mathrm{el},0}$ to only be a function of $Z_i$ (aka $\theta$ below), all force contributions at fixed composition vanish and the second-order energy variation $\propto C^{-1}_{\mathrm{el},0}$ remains configuration-independent. As a result, $C_{\mathrm{el},0}$ does not influence configurational distributions at given $q$ at fixed composition and must thus not be directly integrated into the machine-learning framework.

With configuration- and bias-dependent energetics $E^\alpha(q)$ provided by the RAZOR-MLIP, it is straightforward to derive the desired electronically grand-canonical free energies $\Omega^{Z_i}(\phiE)$ for a defined chemical composition (see section SIII in the SM for more details): First, canonical free energies $F^{Z_i}(q)$ are determined via MD simulation at $q$ and subsequent thermodynamic integration
\begin{equation}
    F^{Z_i}(q) 
    = F^{Z_i}_{0} + \int^{q}_{0} \left.\frac{\partial F^{Z_i}}{\partial q}\right|_{q'} \mathrm{d}q'
    = F^{Z_i}_{0} + \int^{q}_{0} \left\langle\phi^\alpha\right\rangle_{q'} \mathrm{d}q' \quad ,
    \label{eq:TDint}
\end{equation}
where $F^{Z_i}_{0}$ is the free energy contribution at $q=0$, obtained via standard referencing methods \cite{dudzinski_first_2023}, and $\left\langle\phi^\alpha\right\rangle_{q}$ is the average work function of the sampled configurations $\alpha$ of the given interface composition ${Z_i}$ at $q$. In the present work, we approximate
\begin{equation}
    \left\langle\phi^\alpha\right\rangle_{q} \approx 
    \left\langle\phi_0^\alpha\right\rangle_{q} + \frac{q}{C_{\mathrm{el}, 0}} \quad ,
    \label{eq:wf_q}
\end{equation}
where $\left\langle\phi_0^\alpha\right\rangle_{q}$ is the learned work function at $q=0$, and $q/C_{\mathrm{el},0}$ the aforementioned, configuration-independent capacitive response of the implicit solvent model. The transformation to a specified $\phiE$ is then achieved by an appropriate charge-potential mapping $q(\phiE)$, given as the charge that minimizes the "formation free energy" of the interface at $\phiE$ \cite{hormann_electrosorption_2020}
\begin{equation}
    q(\phiE) = \mathrm{argmin}_{q}\left\{F^{Z_i}(q) - q \phiE\right\} \quad .
    \label{eq:q_phi_E}
\end{equation}
The so-determined, optimal formation free energy is identical to the grand potential of the system
\begin{equation}
    \Omega^{Z_i}(\phiE) = \min_{q}\left\{F^{Z_i}(q) - q \phiE\right\} \quad ,
    \label{eq:Omega_phi_E}
\end{equation}
the Legendre transform of the free energy $F^{Z_i}(q)$. Equations (\ref{eq:TDint}) to (\ref{eq:Omega_phi_E}) thus suffice to translate the constant-charge simulations to relevant quantities at constant-potential conditions, i.e. $q(\phiE)$ and $\Omega^{Z_i}(\phiE)$.

We illustrate this workflow and the capabilities of the RAZOR-MLIP approach by considering the potential-dependent adsorption behavior of OH on Cu(100) in aqueous electrolyte, which is of interest in the context of \ce{CO2} reduction at cathodic potentials and surface restructuring at anodic potentials \cite{kunze_situ_2003, simon_potentialdependent_2021, sun_understanding_2022}. The RAZOR-MLIP is based on the NequIP neural network architecture \cite{batzner_e3-equivariant_2022} and the training is performed with density-functional theory (DFT) data generated with Quantum ESPRESSO and the implicit solvent model Quantum ENVIRON \cite{giannozzi_quantum_2009, giannozzi_advanced_2017, andreussi_revised_2012, andreussi_electrostatics_2014}, with full computational and implementation details provided in Section SII of the SM. 

\begin{figure}[t!]
    \centering
    \includegraphics[scale=1]{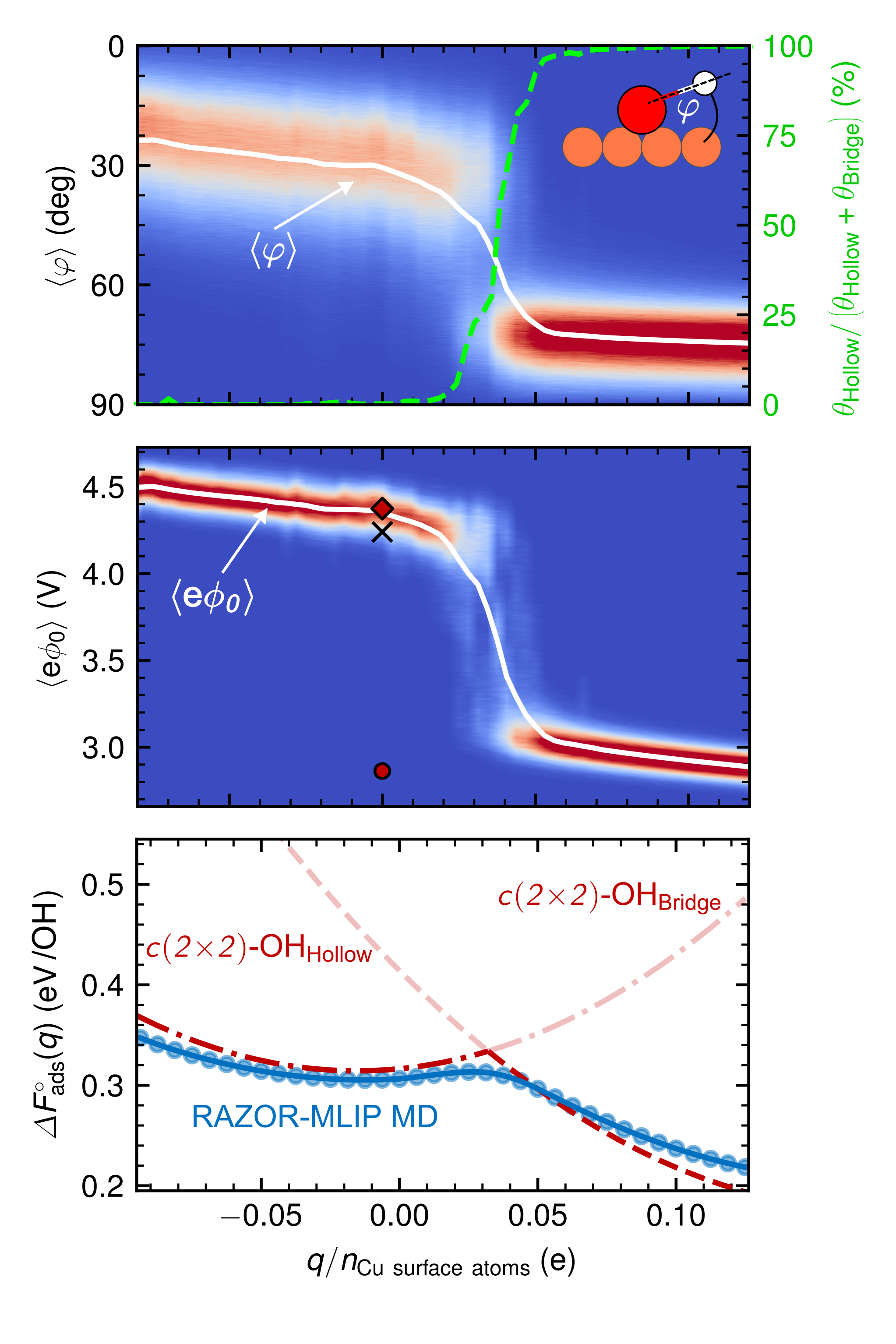}
    \caption{RAZOR-MLIP based MD simulations for a 0.5\,monolayer OH-covered Cu(100) surface at different $q$ (per Cu surface atom). Top panel: Probability distribution of the OH-bond angle $\varphi$ (see inset), together with the average $\langle \varphi \rangle$ (white line) and the fractional coverage of hollow and bridge sites $\theta_{\rm Hollow}/(\theta_{\rm Hollow} + \theta_{\rm Bridge})$ (green line), revealing a charge-induced switch of the adsorption site. Middle panel: Corresponding data for the work function $e\phi_{0}$. Additionally shown are the $e\phi_{0}$ of DFT-optimized, ($q=0$, $T=0$\,K) reference structures: clean Cu(100) (cross), $c(2\times2)$-OH$_{\rm Bridge}$ (diamond), and $c(2\times2)$-OH$_{\rm Hollow}$ (circle). Bottom panel: Constant-charge adsorption free energy $\Delta F^{\circ}_{\rm ads}(q)$ (blue circles), compared to results from DL-corrected \textit{ab initio} thermodynamics for the reference $c(2\times2)$-OH$_{\rm Hollow}$ (red dashed line) and $c(2\times2)$-OH$_{\rm Bridge}$ (red dash dotted line) structures, see text.}
    \label{fig:q_vs_OH_angle}
\end{figure}
$NVT$ MD simulations in supercells with $(4 \times 4)$ surface unit-cells and for $T=300$\,K were run at different $q$ and various OH coverages. With analog results for other coverages reported in the SM, we focus here on the 0.5\,monolayer (ML) case, considered to be the highest possible Cu(100) OH coverage \cite{tiwari_fingerprint_2020}. As summarized in Fig.~\ref{fig:q_vs_OH_angle}, we obtain a charge-induced change of the stable adsorption site. At very negative $q$, OH binds exclusively at the twofold bridge site and lies almost parallel to the surface (average angle $\langle \varphi \rangle \sim 20-30^{\circ}$), whereas at positive $q$ values it prefers the fourfold hollow site with an almost upright orientation (average angle $\langle \varphi \rangle \approx 75^{\circ}$). The switch between the two sites (and the concomitant OH-orientation change) occurs within a narrow charge range $\SI{0.02}{e} \lesssim q \lesssim \SI{0.05}{e}$ per Cu surface atom, and is accompanied by a strong work-function change of $\sim 1.5$\,V. Outside the transition region, charging and temperature effects are small and MD-averaged work functions closely align with those of DFT-optimized, ($q=0$, $T=0$\,K) fully-ordered $c(2 \times 2)$ bridge and hollow reference structures, cf. Fig.~\ref{fig:q_vs_OH_angle}.

Besides the charge-dependent interfacial structures, the RAZOR-MLIP based simulations also give access to important thermodynamic quantities. In Fig.~\ref{fig:q_vs_OH_angle} we include the constant-charge OH adsorption free energies $\Delta F^{\circ}_{\mathrm{ads}}(q)$, computed via the free energy difference (per OH) between the OH-covered and clean Cu(100) slabs as well as the OH standard chemical potential $\mu_{\rm O}^\circ$ (see SM, SIII.3 for details). The comparison to \textit{ab initio} thermodynamics (AITD) \cite{ringe_implicit_2022, rogal_abinitio_2007, hormann_electrosorption_2020} -- the prevalent approximate approach to such energetics at applied charge -- shows excellent agreement (Fig.~\ref{fig:q_vs_OH_angle}~bottom), e.g. for the site switch as derived from the crossing of the free energy curves of the static DFT reference structures. Notably, this agreement prevails over the entire wide range of bias charges, supporting the validity of RAZOR's second-order expansion.

\begin{figure}[htbp!]
    \centering
    \includegraphics[scale=1]{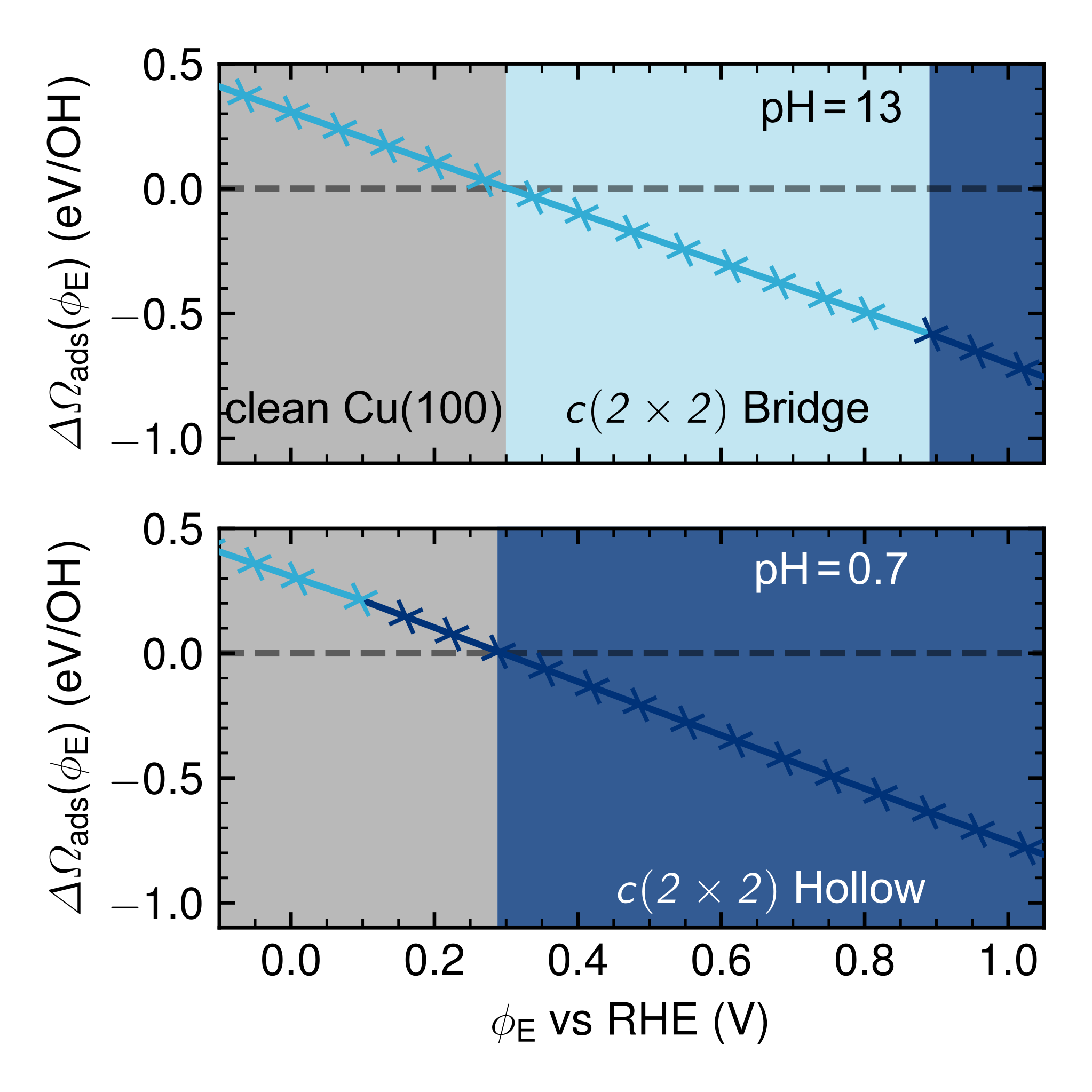}
    \caption{Adsorption free energy $\Delta \Omega_{\rm ads}(\phiE)$ of 0.5\,monolayer OH at Cu(100) as a function of the applied potential $\phiE$ on the reversible hydrogen electrode (RHE) scale. The charge-dependent adsorption site, cf. Fig.~\ref{fig:q_vs_OH_angle}, translates to a potential-induced site switch, with light (dark) blue crosses corresponding to bridge (hollow) site adsorption. This induces a pH-dependence at the onset potentials for OH adsorption above $\phiE \gtrsim 0.3$\,V, i.e. when $\Delta \Omega_{\rm ads}(\phiE)$ becomes negative: At alkaline pH=13 conditions, the preferred adsorption site is bridge (upper panel), whereas at acidic pH=0.7 conditions it is hollow (lower panel).}
    \label{fig:Omega_exc}
\end{figure}
By transforming the constant-charge $\Delta F^{\circ}_{\rm ads}(q)$ to the corresponding constant-potential $\Delta \Omega_{\rm ads}(\phiE)$ we finally reach experimental conditions. Specifically, we employ computational hydrogen electrode (CHE) referencing to the chemical potential \cite{norskov_origin_2004}, $\mu_{\rm O} = \mu_{\rm O}^\circ + k_{\mathrm{B}}T\ln\left(10\right)\cdot\mathrm{pH} + e(\phiE-\phi_\mathrm{SHE})$ and measure $\phiE$ versus the reversible hydrogen electrode (RHE). Figure~\ref{fig:Omega_exc} shows the corresponding results in both acidic and alkaline conditions (see SM for other coverages). Strikingly, this shows that the aforediscussed charge-dependent site switch manifests itself in a pH-dependence of the preferred OH adsorption site: bridge sites in alkaline and hollow sites in acidic conditions, exactly as deduced in experimental studies by Kunze \textit{et al.} \cite{kunze_situ_2003} and Cruickshank \textit{et al.} \cite{cruickshank_situ_1992}.

The agreement between RAZOR-MLIP and AITD results, as well as the consistency with experiment is most gratifying. For one, it provides confidence in the quantitative accuracy of prevalent approximate AITD approaches, at least for systems of intermediate complexity such as the here considered interface with only one adsorbate species on well-defined adsorption sites. On the other hand, it confirms the correctness and robustness of the RAZOR-MLIP-based MD at applied bias, whose true prospective strength lies in the ability to tackle more complex situations, e.g. with mobile adsorbates or explicit interfacial water. In these cases, the structures' dynamical evolution and stabilization at applied potential are crucial and only accessible with significant sampling efforts. Even the present, relatively simple system requires long time scales to reach convergence. As detailed in the SM (Fig. S14), the interfacial transformation dynamics from an initial $c(2\times2)$-OH$_{\rm Bridge}$ to the $c(2\times2)$-OH$_{\rm Hollow}$ upon switch of the charge state can for instance take several hundred picoseconds. Already this is beyond commodity time scales of direct \textit{ab initio} MD simulations, highlighting the importance of fast MLIP-surrogate total energy methods.

In summary, the present "Response Analysis in $z$-ORientation" framework extends existing MLIP approaches to provide computationally efficient and robust access to interfacial atomistic structures and energetics at electrified extended metallic electrodes. Rather than striving to directly machine-learn applied bias conditions, it roots in perturbation theory to capture surface charging effects up to second-order. This reduces the additional learning effort to a MLIP for the work function as the relevant linear response parameter, at concomitant only modest increase in required first-principles training data. RAZOR can be directly implemented with any existing MLIP architecture, which makes it particularly useful in times where general, foundational models become increasingly accurate \cite{deng_chgnet_2023, batatia_foundation_2024, kaplan_foundational_2025, kabylda_molecular_2025, mazitov_petmad_2025}.

Already the first showcase application to the potential-dependent thermodynamics and dynamics of adsorbed OH on Cu(100) reproduces the experimentally reported non-Nernstian pH-dependence of the preferred adsorption site. Yet, this is only a glimpse of the predictive-quality electrochemical modeling capabilities enabled by RAZOR-MLIP acceleration, with future work concentrating on explicitly solvated complex interfacial environments far beyond the reach of direct first-principles based simulations. Specifically, we emphasize the possibility of performing large-scale potentiostated, constant-potential simulations, as well as the use of the local work function decomposition ansatz in developing approximation-free, local descriptions of electrochemical driving forces \cite{Sudarshan_Force2022,Chen_Fundamental2023,beinlich_controlled_2023} or for simulating local work-function variations as probed in Kelvin Probe Force Microscopy \cite{MELITZ20111}.

\bibliography{lib}

\end{document}